\documentclass[aps,prc,showpacs,twocolumn,superscriptaddress,nofootinbib]{revtex4}
\usepackage{amsmath}
\usepackage{amssymb}
\usepackage{graphics}
\newcommand{\etal}{\textit{et al}.}
\begin{document}
\title{Properties of Nucleon Resonances by means of a Genetic Algorithm}
\author{C.~\surname{Fern\'andez-Ram\'{\i}rez}}
\email{cefera@mit.edu}
\affiliation{Center for Theoretical Physics, Laboratory for Nuclear Science and Department of Physics,
Massachusetts Institute of Technology,
77 Massachusetts Ave., Cambridge, MA 02139, USA}
\author{E.~\surname{Moya de Guerra}}
\affiliation{Grupo de F\'{\i}sica Nuclear, Departamento de F\'{\i}sica 
At\'omica, Molecular y Nuclear, Facultad de 
Ciencias F\'{\i}sicas, Universidad Complutense de Madrid, 
Avda. Complutense s/n, E-28040 Madrid, Spain}
\affiliation{Instituto de Estructura de la Materia, 
CSIC, Serrano 123, E-28006 Madrid, Spain}
\author{A.~\surname{Ud\'{\i}as}}
\affiliation{Departamento de Estad\'{\i}stica e Investigaci\'on Operativa,
Escuela Superior de Ciencias Experimentales y Tecnolog\'{\i}a, \\
Universidad Rey Juan Carlos, Camino del Molino s/n, E-28943 Fuenlabrada, Spain}
\author{J.M.~\surname{Ud\'{\i}as}}
\email{jose@nuc2.fis.ucm.es}
\affiliation{Grupo de F\'{\i}sica Nuclear, Departamento de F\'{\i}sica 
At\'omica, Molecular y Nuclear, Facultad de 
Ciencias F\'{\i}sicas, Universidad Complutense de Madrid, 
Avda. Complutense s/n, E-28040 Madrid, Spain}
\date{\today}

\begin{abstract}
We present an optimization scheme that employs a Genetic Algorithm (GA) to determine the properties 
of low-lying nucleon excitations within a realistic 
photo-pion production model based upon an effective Lagrangian. 
We show that with this modern optimization technique it is possible to reliably assess 
the parameters of the resonances and the associated error bars as well as to identify 
weaknesses in the models. 
To illustrate the problems the optimization process may encounter, we provide results 
obtained for the nucleon resonances $\Delta$(1230) and $\Delta$(1700). The former can be easily 
isolated and thus has been studied in depth, while the latter is not as well known experimentally.
\end{abstract}

\maketitle

\section{Introduction}
In recent years, in order to study the properties of low-lying nucleon resonances and
assess their parameters (masses, widths, and electromagnetic coupling constants), significant experimental 
and theoretical efforts have been devoted to the process of meson production from the nucleon, 
which is achieved by exciting the nucleon resonances by means of photonic or electronic probes, 
and to the study of the decays of these resonances into mesons (mainly pions) \cite{Krusche}. 
The parameters of these resonances, predicted by several theoretical models of baryons -- 
lattice Quantum Chromodynamics \cite{Alexandrou},  Skyrme models \cite{Wirzba}, 
quark models \cite{quarks} --  can be compared to the ones extracted from experimental data, 
which usually requires the aid of reaction models. This process of extracting  the 
nucleon excitations parameters from experimental data is thus a crucial requirement in order to validate 
different hadron models, as it provides a guide for improving hadron models and for identifying the most 
reliable ones \cite{PRL}. Together with pion scattering off the nucleon, single pion 
photoproduction is the most suitable process for studying the low-lying baryon spectrum. 
In fact, in recent years the experimental database \cite{SAID} has increased considerably 
and many experimental programs have been run at different facilities such as LEGS (Brookhaven) \cite{LEGS} 
and MAMI (Mainz) \cite{MAMI}.

The extraction of the parameters of the resonances by means of a comparison of the 
results of reaction models to experimental data is an excellent example of a highly involved optimization 
task. 
Problems in which a set of parameters must be established through comparison with experimental 
data are ubiquitous in physics. Often, optimization has been considered a minor topic 
(at times even trivial) by the particle and nuclear physics community which has relied on 
gradient-based optimization tools such as MINUIT \cite{MINUIT}.
Sometimes, however, optimization problems are very complicated and gradient-based routines 
alone are not sufficient for the purpose, because the function to fit presents a complex structure 
with many local optima in which the codes get trapped before reaching anywhere near the desired 
absolute optimum. Thus, until relatively recently, fitting model parameters to data 
has been a kind of art. This was particularly the case when thousands of data needed to be compared 
to the results of sophisticated models that depended on more than just a few parameters. In such cases, 
many instances of the optimization procedure have to be repeated, after manually adjusting 
the parameters, and specific care must be taken to prevent the optimization procedure from getting
stuck at the many possible local minima positions.

Recently, in nuclear and particle physics, more credit is being given to modern optimization 
procedures \cite{ejemplos,Ireland,phd,fernandez06a,Fer07b,Fer07a} and to error estimations 
on the parameters stemming from the fits. Modern and sophisticated optimization 
techniques such as simulated annealing \cite{kirkpatrick} and genetic algorithms (GA) \cite{Goldberg} 
have been developed over the last twenty years and have been applied to problems which are impossible 
to tackle with conventional tools.

In this paper we present a hybrid optimization procedure which combines a GA with a gradient-based 
(''hill-climbing'') routine \texttt{E04FCF} from the NAG library \cite{NAG}. The GA performs the bulk 
of the optimization efforts, ensuring that the parameter space is fully surveyed and local minima 
are avoided, while the conventional gradient-based routine, when applied to the preliminary minima 
found by the GA, provides fine-tuning and speeds up convergence. We have applied this tool to a 
complex, multi-parametric optimization problem, namely the determination of nucleon 
resonances parameters by comparing the results of a realistic model for the photo-pion 
production reaction to data. As a by-product, the optimization procedure provides insight into the reliability 
of the values (error bars) of the parameters extracted and information on their physical significance.

This paper is organized as follows: in Section \ref{sec:reaction_model} we briefly present the model 
for pion photoproduction on free nucleons from threshold up to 1.2 GeV developed in 
Refs. \cite{phd,fernandez06a,Fer07b}. In Section \ref{sec:strategy} we present the strategy 
applied to solve the problem. In Section \ref{sec:genetic-algorithm} we present the GA in detail. 
In Section \ref{sec:results} we show the results obtained by the algorithm, analyze its performance 
and comment on the error bar estimates and the physical significance of the parameters extracted. 
Finally, in Section \ref{sec:conclusions} we present our conclusions.

\section{The reaction model}\label{sec:reaction_model} The reaction model is based upon a phenomenological 
Lagrangian and it allows us to isolate the contribution of the resonances, calculate their bare properties, 
and compare these properties with the values provided by nucleonic models \cite{phd,fernandez06b}. In addition 
to Born terms (those which involve only photons,  nucleons and pions) and vector-meson exchange terms 
($\rho$ and $\omega$), the model includes all four star resonances quoted in the Particle Data Group 
(PDG) \cite{PDG2006} up to 1.8 GeV of mass and up to spin-3/2: $\Delta$(1232), N(1440), N(1520), $\Delta$(1620),
N(1650), $\Delta$(1700), and N(1720). The internal structure of the nucleonic excitations shows up in the 
values of the electromagnetic coupling constants that appear in the Lagrangian. The model displays 
chiral symmetry, gauge invariance, and crossing symmetry, and incorporates a consistent treatment of the 
interaction with spin-3/2 particles that avoids well-known pathologies of previous 
models \cite{phd,fernandez06a,Pascalutsa}.
Furthermore, the dressing of the resonances is considered by means of a phenomenological width which takes 
into account decays into one and two $\pi$'s and one $\eta$. This width is included in a way that 
fulfils crossing symmetry and thus it contributes to both the direct and crossed channels of the resonances.
We assume that the final state interactions (FSI) in the $\pi N$ rescattering factorize and can be 
included through the distortion of the $\pi N$ final state wave function. We include this distortion in a 
phenomenological way by incorporating a phase $\delta_{\text{FSI}}$ to the electromagnetic multipoles. We fix 
this phase so that the total phase of the electromagnetic multipole is identical to that of the energy 
dependent solution of SAID \cite{SAID}. In this way, we disentangle the parameters of the electromagnetic 
vertex from the FSI effects.

\section{Minimization strategy}\label{sec:strategy} Our minimization procedure follows  the one 
in \cite{Ireland} although we use a more sophisticated GA and employ the \texttt{E04FCF} routine from the NAG 
library \cite{NAG} instead of \texttt{MINUIT} \cite{MINUIT} code.  We apply the minimization scheme to a 
realistic meson production model and the aim of our minimization is different. While in \cite{Ireland} the 
aim was to establish the existence of certain resonances, in this paper our goal is to determine the 
parameters of well-established nucleon resonances and to obtain estimates on the reliability of these 
parameters and their associated error bars.

The function to minimize is the $\chi^2$ defined by
\begin{equation}
\chi^2 = \sum_j \frac{\left( \mathcal{M}^{exp}_j-
\mathcal{M}^{th}_j \left( \lambda_1, \dots, \lambda_n \right) 
\right)^2}{\left( \Delta \mathcal{M}^{exp}_j \right)^2} ,
\end{equation}
where ${\mathcal M}^{exp}$ stands for the current energy independent extraction of the multipole analysis of 
SAID up to 1.2 GeV for $E_{0+}$, $M_ {1-}$, $E_{1+}$, $M_{1+}$, $E_{2-}$, and $M_{2-}$ multipoles in the three 
isospin channels $I=\frac{3}{2},p,n$ for the $\gamma p \to \pi^0 p$  process \cite{SAID}.
$\Delta \mathcal{M}^{exp}$ is the experimental error and $\mathcal{M}^{th}$ is the multipole value given by the 
model. It depends on parameters $\lambda_1$, $\dots$,$\lambda_n$. We have taken into account 1,880 data for the 
real part of the multipoles and the same amount for the imaginary part.  Thus, 3,760 data points have been used 
in the fits. Unlike cross-sections or 
asymmetries, electromagnetic multipoles are not directly measured quantities and some elaboration of 
the raw experimental data is needed to obtain these multipoles.  However, we 
have chosen, as it is very often done in this field, to fit electromagnetic multipoles instead of other 
observables. Several reasons are mentioned when fitting to multipoles. On one hand, 
electromagnetic multipoles are more sensitive to coupling properties than other observables, so deficiencies 
in the model may show up more clearly. The second reason is that, in principle, all the observables can 
be expressed in terms of multipoles. Thus, if the multipoles are properly fitted by the model, so should be 
other observables. 

In order to determine the resonance parameters that best fit the data, 
we have written a hybrid optimization code based on a GA combined with the 
\texttt{E04FCF} routine from the NAG library \cite{NAG}. Although GA, are computationally more expensive  
than other algorithms, in a minimization problem it is much less likely for them to get stuck at 
local minima than it is for other methods, namely gradient-based minimization methods. GAs allow us to explore 
a large parameter space more efficiently. Thus, in a multi-parameter minimization such as the one we face here, they 
are probably a very efficient way of searching for the best minimum. In the next section we will go through 
the details of the GA.

The parameters for the model ($\lambda_j$) are divided into two different kinds:
(i) Those that are obtained from models or experiments other than pion photoproduction, 
namely vector-meson coupling constants (three parameters) and masses and widths of the nucleon resonances 
(fourteen parameters, one mass and one width for each resonance which have been taken from \cite{Vrana}), and
 (ii) those that are extracted from pion photoproduction data, namely electromagnetic coupling constants 
(fifteen parameters) and the cutoff $\Lambda$ for Born terms and vector-meson exchanges. We have allowed 
the algorithm to vary all the parameters (see Tables \ref{tab:parameters1} and \ref{tab:parameters2}). However, the parameters in the first group have been varied within a 
very small range, the experimentally allowed values for the vector-meson coupling constants 
and $\pm 2$ MeV for the masses and widths of the nucleon resonances. The reason for allowing these parameters 
to vary, even though  the range of variation is minimal, is to make room for the algorithm to search for 
the global minimum and to take into account the error bars for these parameters into the possible solution. 
This should help to prevent the algorithm for being trapped in local minima.

The variation range for the second group of parameters are chosen to explore a large region of parameter 
space. Hence we avoid introducing prejudgments on their values based on previous analysis. We prefer to use 
the helicity amplitudes (for their definition and connection with coupling constants see 
Refs. \cite{phd,fernandez06a,PDG2006}) to define the ranges, instead of the electromagnetic coupling 
constants. We allow them to vary in the range $\left[ -1,1 \right]$ GeV$^{-1/2}$.

The cut-off $\Lambda$ is included in the form factors that multiply the Born terms and vector-meson exchange 
invariant amplitudes. We use the form factors suggested in \cite{Davidson01-1}, which respect gauge invariance 
and crossing symmetry. For these Born terms
\begin{equation}
\begin{split}
\hat{F}_B(s,u,t)= &F(s)+F(u)+G(t)-F(s)F(u) \\
&- F(s)G(t)-F(u)G(t)+F(s)F(u)G(t) ,
\end{split}
\end{equation}
where
\begin{eqnarray}
F(l)&=& \left[1+ \left( l-M^2 \right)^2/\Lambda^4 \right]^{-1}, 
\quad l=s,u ;\\
G(t)&=& \left[1+ \left( t-m_\pi^2 \right)^2/\Lambda^4 \right]^{-1} .
\end{eqnarray}
and $s$, $u$, and $t$ are the Mandelstam variables.
For vector mesons, we adopt $\hat{F}_V(t) = G(t)$ with the change $m_\pi \to m_V$. 
To reduce the number of free parameters for the model we use the same $\Lambda$ for both vector mesons and Born 
terms.

The form factors take non-resolved structure effects and higher order terms in the scattering matrix 
expansion into account. Thus, the cut-off $\Lambda$ is related to the energy scale of the effective theory and 
the sensible values
 for $\Lambda$ should be of the order of the nucleon mass (actually, in our best fit we obtain $\Lambda=0.943$ 
MeV). For this reason, in the minimization process, we restrict $\Lambda$ to the range $\left[ 0.1,2.0 \right]$ 
GeV.

\begin{figure}
\rotatebox{-90}{\scalebox{0.32}[0.32]{\includegraphics{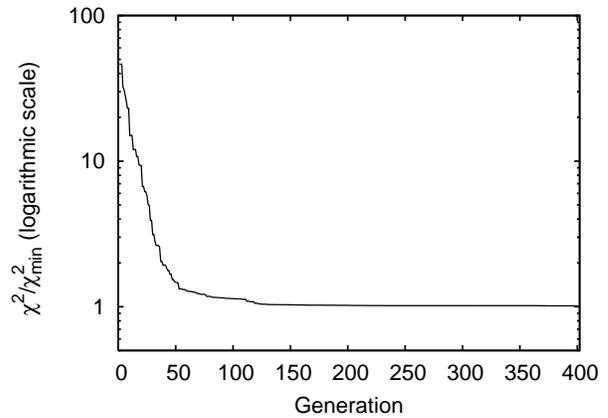}}}
\caption{Example of the evolution for a champion in one run of the GA. For the first generations 
(up to generation 40 or 50) evolution is driven by crossover. After this, small improvements are seen due 
to mutations.}
\label{fig:minimization_evolution}
\end{figure}

In order to perform the minimization, the range of variation of each parameter is mapped into the $[0,1]$ 
interval for the GA and into $(-\infty , +\infty)$ for the \texttt{E04FCF} routine. This latter step is done 
by means of the transformation
\begin{equation}
x_j=\arcsin \left[ \frac{\lambda_j-\lambda^{max}_j}{\lambda^{max}_j
-\lambda^{min}_j-1} \right] ,
\end{equation}
where
$\lambda_j$ is the model parameter, $x_j$ is the mapping of $\lambda_j$
into $(-\infty , +\infty)$,
$\lambda^{max}_j$ is the highest value of the range of variation, and $\lambda^{min}_j$ is the lowest value.  
With regard to the range of variation allowed for the parameters, we must note that gradient routines work more 
efficiently if variations of similar magnitude on each of the search parameters introduce a similar variation 
on the function to minimize. The \texttt{E04FCF} user is advised to explore the region of parameters to 
minimize and to provide adequate rescaling of the problem before calling the routine. While the NAG library 
provides tools that help in this task, in our combined algorithm we take advantage of the knowledge obtained 
on the variation of the objective function during the previous evaluations performed by the GA.
We use this exploration to normalize the $\chi^2$ to unity and to rescale all the parameters affecting this 
function so that, according to the last evaluations of the best individuals explored by the GA, after rescaling 
of both the parameters and the function to optimize, the region explored by the NAG \texttt{E04FCF} routine 
in its search for the minima is expected to lie in a hypercube of unit volume. We have indeed verified 
that this normalization and rescaling procedure improves NAG routine performance.

\begin{table}
\caption{Ranges for the parameter values of the nucleon resonances. Masses and decay widths have been 
taken within the ranges provided by \cite{Vrana}.
The helicity amplitudes are denoted by $A^I_\lambda$, where $I$ stands for isospin and $\lambda$ for the 
helicity of the initial photon-nucleon state.} \label{tab:parameters1}
\begin{ruledtabular}
\begin{tabular}{lccc}

&$M^*$ (GeV)& $\Gamma$ (GeV)&  $A^I_{\lambda}$ (GeV$^{-1/2}$)\\

\hline
$\Delta$(1232) & [1.215,1.219] & [0.094,0.098] & [-1,1] \\
N(1440)             & [1.381,1.385] & [0.314,0.318]  & [-1,1] \\
N(1520)             & [1.502,1.506] & [0.110,0.114]  & [-1,1] \\
N(1535)             & [1.523,1.527] & [0.100,0.104]  & [-1,1] \\
$\Delta$(1620) & [1.605,1.609] & [0.146,0.150]  & [-1,1] \\
N(1650)             & [1.661,1.665] & [0.238,0.242]  & [-1,1] \\
$\Delta$(1700) & [1.724,1.728] & [0.116,0.120]  & [-1,1] \\
N(1720)             & [1.740,1.755] & [0.119,0.278]  & [-1,1] \\
\end{tabular}
\end{ruledtabular}
\end{table}

\begin{table}
\caption{Ranges for the values of the parameters of vector mesons 
and cut-off $\Lambda$.} \label{tab:parameters2}
\begin{ruledtabular}
\begin{tabular}{lccc}

 $F_{\omega NN}$  &  $[20.61, 21.11]$  \\
 $K_\omega$           &  $[-0.17,-0.15]$     \\
$K_\rho$                   &  $[6.1,6.3]$    \\
$\Lambda$  (GeV)  &  $[0.1, 2.0]$   \\
\end{tabular}
\end{ruledtabular}
\end{table}

\begin{figure}
\rotatebox{0}{\scalebox{0.6}[0.6]{\includegraphics{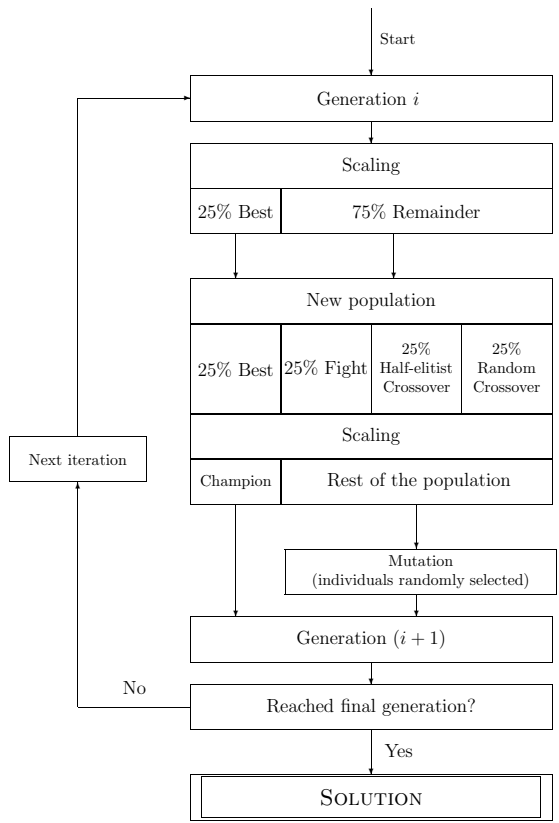}}}
\caption{GA scheme (see text in section 
\ref{sec:genetic-algorithm}).}
\label{fig:genetic_work}
\end{figure}

Our minimization strategy includes the following aspects:
\begin{enumerate}
\item A first generation is made out of individuals randomly generated within reasonable values of the parameters. 
\item Next, the GA is run for 400 \textit{generations} (see definition further on). This number is determined 
after 
inspecting the best individual evolution for each generation and from comparisons with benchmark 
problems of similar size. We do not really need that many as 400 generations 
(see Fig. \ref{fig:minimization_evolution}), but we preferred to let the algorithm run for more generations 
than necessary in order to ensure that convergence was achieved.
\item After the 400 generations have been run, we introduce the GA solution as the initial value for 
the \texttt{E04FCF} routine from NAG libraries \cite{NAG}. We use the routine for fine-tuning. 
The \texttt{E04FCF} routine implements an algorithm that looks for the unconstrained minimum of a sum 
of squares
\begin{equation}
\text{Minimize} \Big{[} F(x_1, \dots, x_n)  =
\sum^m_{j=1} |f_j \left( x_1, \dots, x_n\right)|^2 \Big{]},
\end{equation} 
of $m$ nonlinear functions in $n$ variables ($m \geq n$).
This algorithm does not require the derivatives to be known.
From a starting point $x^{(1)}_1$, $\dots$, $x^{(1)}_n$ (in our case supplied by the GA) the routine applies a 
Quasi-Newton method in order to find the minimum.
This method uses a finite-difference approximation to the Hessian matrix to define the search direction. It 
is a very accurate and fast converging algorithm once we have an initial solution that is close to the minimum 
we seek. Therefore, it is well suited for our fine-tuning purpose.

We note that many attempts to solve our optimization procedure solely by means of \texttt{E04FCF}
completely failed, even when we guided the initial ranges of the parameters by hand. The NAG routine got 
stuck in the first local minimum, usually very far from the one obtained by the GA.

\item We store the solution obtained by the combined algorithm and we start again, by generating a 
different random seed for the initial population of the GA. After running the minimization code twenty
times, we obtain twenty different minima. If we find that all the $\chi^2$ divided by $\chi^2_{min}$  
(the minimum $\chi^2$ among all the fits) are close  to unity, we stop the fitting procedure.
\end{enumerate}

\section{Genetic algorithm}\label{sec:genetic-algorithm}

\begin{figure}
\rotatebox{0}{\scalebox{0.4}[0.45]{\includegraphics{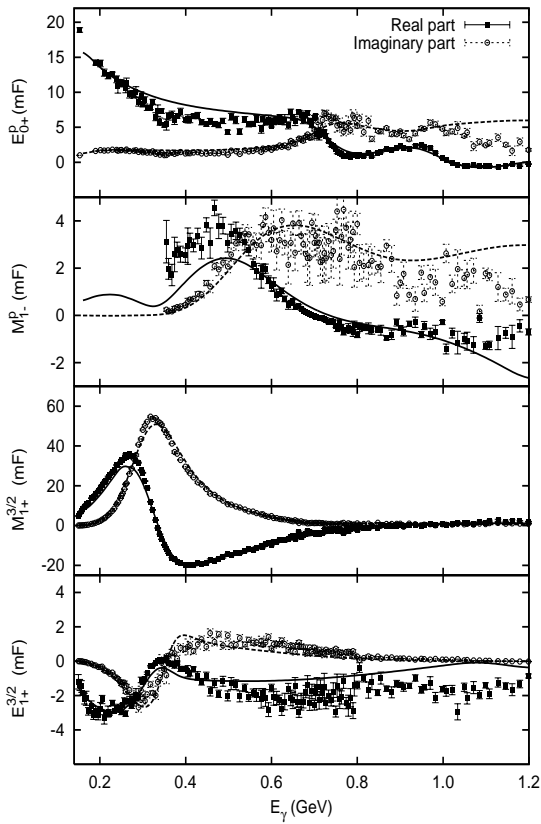}}}
\caption{Examples of the fits obtained to the electromagnetic multipoles 
for the reaction $\gamma p \to \pi^0 p$. Curve conventions: Solid: Real part of the electromagnetic 
multipole; Dashed: Imaginary part of the electromagnetic multipole. Data are from Ref. \cite{SAID}.} 
\label{fig:fig_multipoles}
\end{figure}

Genetic Algorithms are a specific kind of stochastic optimization methods based upon the idea of evolution. 
There are many excellent textbooks on GA \cite{Goldberg}. Here we will describe the main features of GA that 
are needed to understand our implementation. GAs encode the possible solutions to the proposed 
problem and deal with many of these solutions at the same time. Indeed, a set of these possible solutions 
(also called individuals or genes) form a population. Each individual in the population is classified 
according to its fitness value, computed in terms of some objective function related to our 
optimization problem. In our case, the individual encodes the parameters of the Lagrangian and the objective 
function is essentially the $\chi^2$ of the multipole values compared to the prediction of each Lagrangian 
represented in the population. GAs implement operators such as crossing among individuals and 
mutation \cite{Eshelman91}. As long as both the encoding of the problem and the GA operators exhibit 
good schema properties (that means that the offspring obtained after breeding two or more individuals 
with some good properties in terms of fitness are, more often than not, more fit than any of their parents), 
the \textit{evolution} or repeated application of the genetic operators on the population, combined with a 
mechanism of natural selection (survival of the fittest), would cause some individuals to accumulate 
the good properties (sub-schema) initially distributed among different individuals in the early population. 
Provided that the number of individuals in the population is large enough for many good sub-schemas to be 
represented in at least some individual of the population, then the GAs would evolve toward very fit 
individuals, that is, good solutions to the problem. In this work, the obvious sub-schema are the parameters 
of each resonance and a simple encoding in which every individual is composed of a set of possible values 
for the parameters of our Lagrangian, would do the job. We encode the possible solutions to the 
problem (i.e., a complete set of parameters for the Lagrangian) as a series of integer numbers within 
the range from 0 to a maximum value $N$. For each parameter, this integer number represents the value of said 
parameter within the range desired by the user. For instance, a stored value of 0 would indicate that the value 
of the corresponding parameter equals the minimum allowed within the range. Conversely, the maximum value $N$ 
would represent the stored parameter reaching the maximum allowed within the range. We denote this 
maximum value of these integers $N$ as the \textit{granularity} of our encoding strategy. A large value of $N$ 
implies a very thin granularity, that is, relatively small changes in each parameter are 
possible in our encoding strategy and individuals that are very similar in terms of the parameters they represent and 
consequently, in their fitness, can be encoded. On the other hand, if we want to sample the parameter space 
with reasonable density, a too thin granularity would require a very large number of individuals. 
As we have just mentioned, an important choice to make for every GA is the number of individuals in the 
population.  When the population size increases, the chances for relatively less fit individuals of mating 
with other individuals and generating better offspring before disappearing from the population, decrease 
exponentially. We must realize that even the less fit individuals (some of them) may have good 
sub-schema needed to encoded the best solution. Some of these sub-schema may not be present in 
other more fit individuals in the population, at least during the early stages of the evolution. According to 
results of tests with our Lagrangian as well as benchmarks with other functions that are easy to compute and 
have well known minima, we have determined that the maximum number of individuals we may safely employ in a 
population for our GA is around 400. For this size of the population, granularity values from 100 to 1000 
have been employed in our GA without problems.

In what remains of this section we simply provide a detailed explanation on how the GA we have programed works. 
Our GA proceeds as follows \ref{fig:genetic_work}:

\begin{enumerate}
\item Initial population. We provide a first {\em generation} consisting of individuals (400 in our calculations) 
that are randomly generated with reasonable values of the parameters \cite{Goldberg85}.

\item Selection scheme. The genetic algorithm we use employs a scaled selection scheme and employs the 
elitist model \cite{deJong}. In this model, the best individual (or champion) from the previous generation is 
always included in the current population, ensuring that the best solution this far is preserved. This decreases 
significantly the time the GA takes to find an acceptable solution. It has been proved \cite{deJong} that the GA 
which introduces elitism (that is, the guaranteed survival of the champion at every step of the GA evolution) 
will eventually converge to the absolute optimum, while, in general, the ones that do not protect the champion 
will never reach the optimum \cite{Rudolph}.

With regard to the remainder of the population, besides the champion, the individuals from the previous 
generation (that is, the population in its earlier state) are ranked according to the \textit{fitness} function, 
in our case the $\chi^2$ value. After this step, we introduce \textit{scaling} of the 
population \cite{goldberg2} determining the probability that an individual has to mate and survive. We provide 
a 0.8 probability to the worst individual and 1.0 to the best one. This is done in order to maintain genetic 
diversity. Indeed, it is necessary to prevent that the best and the worst individuals have a too different 
survival probability. If we do not take care to preserve genetic diversity in this way, the appearance of a 
very fit individual would make the forthcoming offspring collapse to the characteristics of that 
particularly fit individual too soon. Another important technique to maintain diversity is \textit{mutation}, 
which is discussed further on.

\item After scaling, we classify the population into two sets. Set (a) is composed of the best 25\% 
of the individuals and set (b) by the remaining 75\%. We produce the new generation in the following way:
\begin{itemize}
\item 25\% of the individuals are taken from the most fit ones from the previous generation. 
That is, set (a) is copied into the next generation.
\item Another 25\% is selected through a fight among all the individuals (tournament). 
The outcome of the fight is randomly decided, depending on probability. Even in the least favorable case 
(that is, if the worse individual fights with the best one), the winning probability of (the worst) individual 
is 15\%. Winning probabilities are computed accordingly to the fitness of each contender.
\item Another 25\% is obtained by means of half-elitist crossover. This means that we mate an individual from 
the best 25\% of the previous generation (set (a)) with any other individual in sets (a) or (b). Both 
individuals are picked randomly from their respective sets.
\item The remaining 25\% of the offspring are generated by mating individuals that are selected randomly without 
restrictions from sets (a) or (b).
\end{itemize}

We apply two different kinds of crossover: \textit{one point} crossover and 
\textit{arithmetic} crossover \cite{goldberg2}. In one point crossover, a random crossover point for both 
parents is selected. We split each chromosome from the parents into two pieces. We take the second piece of the 
second parent and attach it to the first piece of the first parent. In this way we obtain an individual 
that is a mixture of the two original ones. For the arithmetic crossover, we choose at random a number $r$ 
between 0 and 1, and the offspring is calculated weighting the parents with weight $r$ and $(1-r)$.
\begin{equation}
\lambda_i^{\text{offspring}}=r \cdot \lambda_i^{\text{parent 1}}+
(1-r) \cdot \lambda_i^{\text{parent  2}}
\end{equation}
Half of the crossovers our GA implements are one point and the other half are arithmetic.  The kind 
of crossover to apply to a given pair of parents is chosen at random.

\item We evaluate the new population and identify the new champion. As previously mentioned, it will be 
preserved (elitism). We select other individuals to mutate from the rest of the population excluding the champion. 
Indeed, in each iteration of our GA we introduce as many mutations as the number of individuals in the 
population divided by three. These mutations are distributed at random among all the individuals 
(excluding the champion) of the population generated following the previous steps.
We apply two types of mutation \cite{Spears}. The \textit{permutation} mutation exchanges two parameters 
selected at random. The \textit{gaussian} mutation changes the value of a parameter by a small amount.  
The amount of change induced by this mutation is random within a small range.
The reason to introduce mutations is that, quite often, the crossover operator and the selection method 
are too effective and they end up driving the GA toward a population of individuals that are almost exactly 
the same. When the population consists of similar individuals, the likelihood of finding new solutions typically 
decreases. The mutation operator introduces an additional randomness into the search. It helps to maintain 
diversity and to find solutions that crossover alone might not discover.

\item After these steps are taken, we say that a new generation is built. If we have not reached the limit 
in the number of generations, we run the algorithm again with the current set of individuals as the initial 
population.

When the maximum number of generations has been reached, we take the set of parameters encoded by the champion 
as the solution given by GA to our problem. If sufficient generations have been run, most of the individuals 
will have close values for the fitness function.
\end{enumerate}

\begin{table}
\caption{Helicity amplitudes obtained in the fits in GeV$^{-1/2}$.} \label{tab:helicities}
\begin{ruledtabular}
\begin{tabular}{lcccccc}

&$A^p_{1/2}$ & $A^\Delta_{1/2}$& $A^n_{1/2}$ & $A^p_{3/2}$ & $ A^\Delta_{3/2}$& $A^n_{3/2}$\\

\hline
$\Delta$(1232) & -- & $-0.120$& -- & -- & $-0.229$& -- \\
N(1440)             &$0.060$& -- & $-0.089$& -- & -- & -- \\
N(1520)             &$-0.007$& -- &$0.032$&$0.107$& -- &$-0.085$ \\
N(1535)             &$0.014$& -- &$-0.137$& -- & -- & -- \\
$\Delta$(1620) & -- &$-0.023$& -- & -- & -- & -- \\
N(1650)             &$-0.022$& -- &$0.003$& -- & -- & -- \\
$\Delta$(1700) & -- &$0.139$& -- & -- &$-0.127$& -- \\
N(1720)             &$0.143$& -- &$0.126$&$-0.004$& -- &$-0.444$ \\
\end{tabular}
\end{ruledtabular}
\end{table}

 It has been proven that there is no optimal algorithm that adapts well (that is, reaches a solution in 
the least number of evaluations) to all kind of problems. This is often referred to as the 
\textit{no free lunch theorem} in optimization \cite{wolpert}.
Our goal here however is not to find the optimal algorithm that obtains the minimum to our problem in less 
evaluations but, rather, to develop a general tool that can be applied to many different models of parameter 
data fitting without specific fine-tuning nor human intervention, even if the performance of the tool is 
sub-optimal in terms of the number of operations.  In this regard, GAs are a handy choice, as they are 
suitable for 
many different problems. Thanks to scaling and elitism, our GA converges neither too quickly nor too slowly 
and generally it is able to find good candidates for the global optimum. 

When the individuals are very fit, it can be hard for the GA to evolve further, mainly because the path 
to the best individual may involve two or more consecutive mutations where each of these mutations on their own 
will produce a less fit individual that will sooner be removed from the population. The occurrence of such two 
favorable mutations in the same individual is unlikely and tailored procedures must be implemented to introduce 
specific mutations that are adequate for particular problems, or more complex operators like the 'tunneling 
algorithm' \cite{Montalvo} or complex rules to encode the values of the functions, like Mendelian operators 
implementing a non-dominant character for some genes \cite{Song}. In our work, however, we prefer to employ a 
hybrid optimization method that combines a standard hill-climbing algorithm with a GA. Hybrid optimization 
methods have been under study intensively \cite{Ireland,goldberg3}. We have compared several ways of hybridizing 
GAs and conventional gradient based hill-climbing algorithms, such as introducing the hill-climbing algorithm
 as another mutation operator. However, we have noticed that this will only make the GA converge sooner, very 
often too soon, resulting in it getting stuck at any of the many local minima.
From our experience, if the hill-climbing procedure is introduced just at the end of the evolution, when 
the GA has converged, the best results are achieved and a robust algorithm that requires no human intervention 
is this way configured. Also, no granularity is introduced in this final step of optimization. Indeed,
the NAG routine is not restricted to integer values of the parameters, but instead represents each parameter as 
floating point values. Thus, we can also consider that the GA finds the best optimum that can be represented 
within the grid implied by the granularity $N$, and starting from this point of the grid, the NAG routine 
refines a search not bound to any grid values.

\section{Results}\label{sec:results}

\begin{figure}
\rotatebox{-90}{\scalebox{0.32}[0.32]{\includegraphics{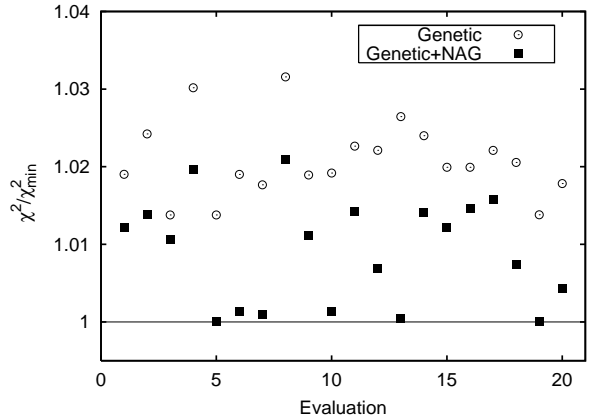}}}
\caption{Many local minima and the effect of 
the fine tuning performed by the \texttt{E04FCF} routine in the 
$\chi^2/\chi^2_{min}$ are shown. Conventions: Open circles, $\chi^2/\chi^2_{min}$ obtained by the GA alone 
(400 generations with 300 individuals each); 
Solid squares: $\chi^2/\chi^2_{min}$ obtained by the GA plus the NAG routine.} 
\label{fig:finetunning}
\end{figure}

In Fig. \ref{fig:fig_multipoles} we show examples of fits to electromagnetic multipoles for the 
$\gamma p \to \pi^0 p$ process and the overall agreement obtained. The values of the parameters are summarized in Table \ref{tab:helicities}. 
In Fig. \ref{fig:minimization_evolution} we display an example of the evolution of the champion along 
the generations. Two hundred generations are sufficient enough to achieve convergence, but we run the 
algorithm for another two hundred generations to see the effects of mutations, which can reach areas 
of the parameter space that are not being fully surveyed by means of crossover.

\begin{figure*}
\rotatebox{-90}{\scalebox{0.55}[0.55]{\includegraphics{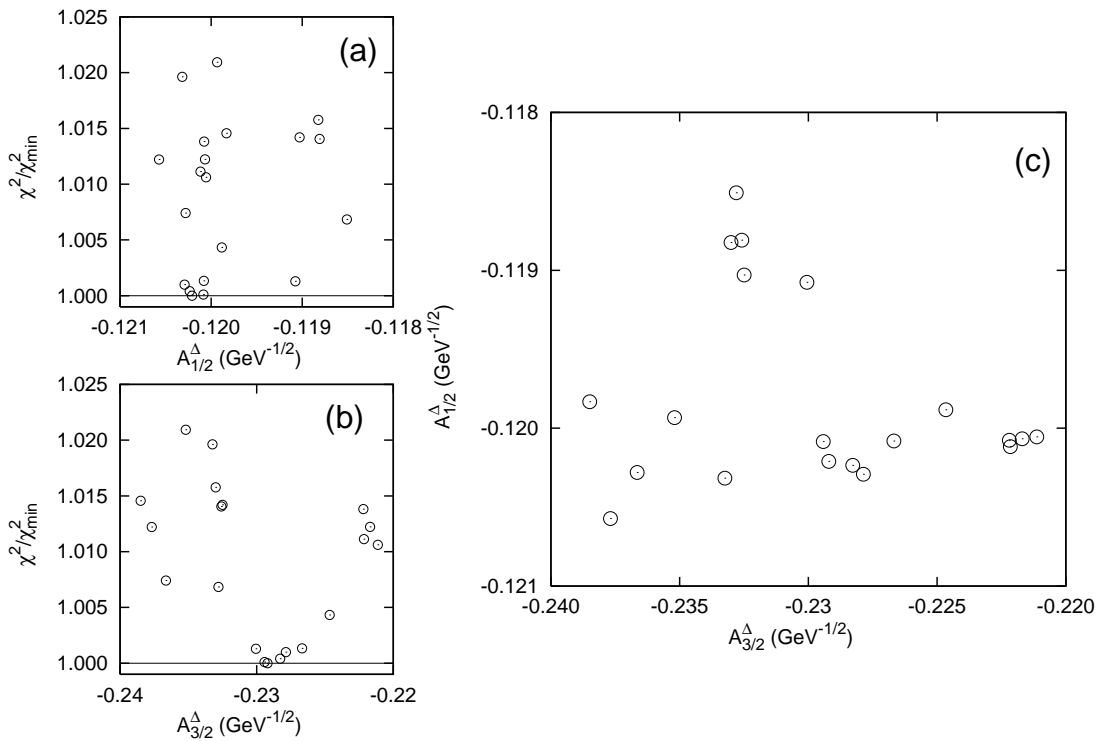}}}
\caption{Helicity amplitudes (equivalent to the coupling constants of the Lagrangians) of the $\Delta$(1232).  
In all the figures we show the twenty minima obtained in the
full minimization procedure (GA$+$NAG, see Fig. \ref{fig:finetunning}).
The upper-left hand figure (a) shows the $\chi^2/\chi^2_{\text{min}}$ versus the amplitude 
$A_{1/2}^\Delta$. The lower-left hand figure (b) shows the $\chi^2/\chi^2 
{\text{min}}$ versus the amplitude $A_{3/2}^\Delta$. 
The right panel (c) shows $A^\Delta_{1/2}$ versus $A^\Delta_{3/2}$ parameters.} 
\label{fig:delta1232}
\end{figure*}

We observe that at the early stages of the evolution the fitness function improves quickly, as crossover works to 
concentrate the good schema from other individuals into a good individual. Actually, a very steep slope 
in this region might indicate that evolution is too fast and that less fit individuals could 
disappear from the population before their good properties are transmitted to more fit individuals.

When a jump in the $\chi^2/\chi^2_{\text{min}}$ happens, it is due to the appearance of a more fit new 
individual, either due to crossover or to mutation.
In Fig. \ref{fig:finetunning} we can verify the existence of many local minima (so this is certainly an 
ill-posed optimization problem) and the fine tuning achieved by the NAG routine which improves minima by 
approximately 2\%.

\begin{figure*}
\rotatebox{-90}{\scalebox{0.55}[0.55]{\includegraphics{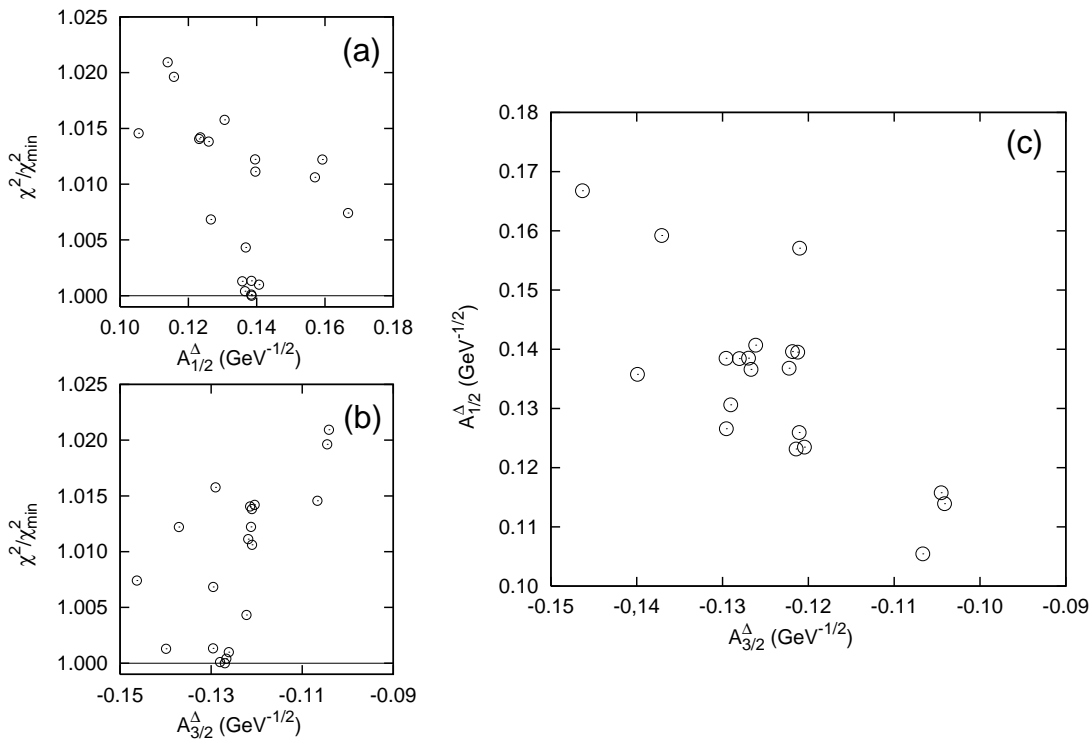}}}
\caption{Helicity amplitudes (equivalent to the coupling 
constants of the Lagrangians) of the $\Delta$(1700). 
In all the figures we show the twenty minima obtained after the full minimization procedure 
(GA$+$NAG see Fig. \ref{fig:finetunning}).
The upper-left hand figure (a) shows the $\chi^2/\chi^2_{\text{min}}$ versus the 
amplitude $A_{1/2}^\Delta$. The lower-left hand figure (b) shows the $\chi^2/\chi^2_{\text{min}}$ versus 
the amplitude $A_{1/2}^\Delta$. The right figure (c) shows $A^\Delta_{1/2}$ versus $A^\Delta_{3/2}$.} 
\label{fig:delta1700}
\end{figure*}

An important issue to consider in GAs is efficiency.  As we have already mentioned, the parameter space has to be 
discretized with a certain granularity and the algorithm searches for the best solution within the discretized 
version of the parameter space. The size of this space significantly affects the efficiency of the algorithm, 
thus a balance between granularity and computing time has to be achieved. The gradient based routine allows us 
to gain 
precision and efficiency because we do not need the GA to reach the minimum, we simply need it to provide a 
value 
close enough for the \texttt{E04FCF} routine can reach it. In other words the GA has to reach the region where 
the minimum lies, and once in this region, reaching the minimum is a task for the gradient-based routine.

We must emphasize that the use of our algorithm is unattended. That is, we submit the script that starts 20 
instances of the GA$+$NAG procedure, and after the equivalent to five CPU-days (Opteron, 2 GHz), we get the 
results for the optimized set of parameters. No further human intervention was needed to choose initial values of 
the parameters or to guide the evolution. While the GA$+$NAG may require more (costly) evaluations of the 
objective function, it is robust and needs no training nor good guesses of the initial parameters. Now that 
computer power seems to be an increasingly available resource, the unattended mode of operation makes this 
hybrid algorithm a very interesting alternative for these optimization problems.

\begin{figure}
\rotatebox{0}{\scalebox{0.45}[0.45]{\includegraphics{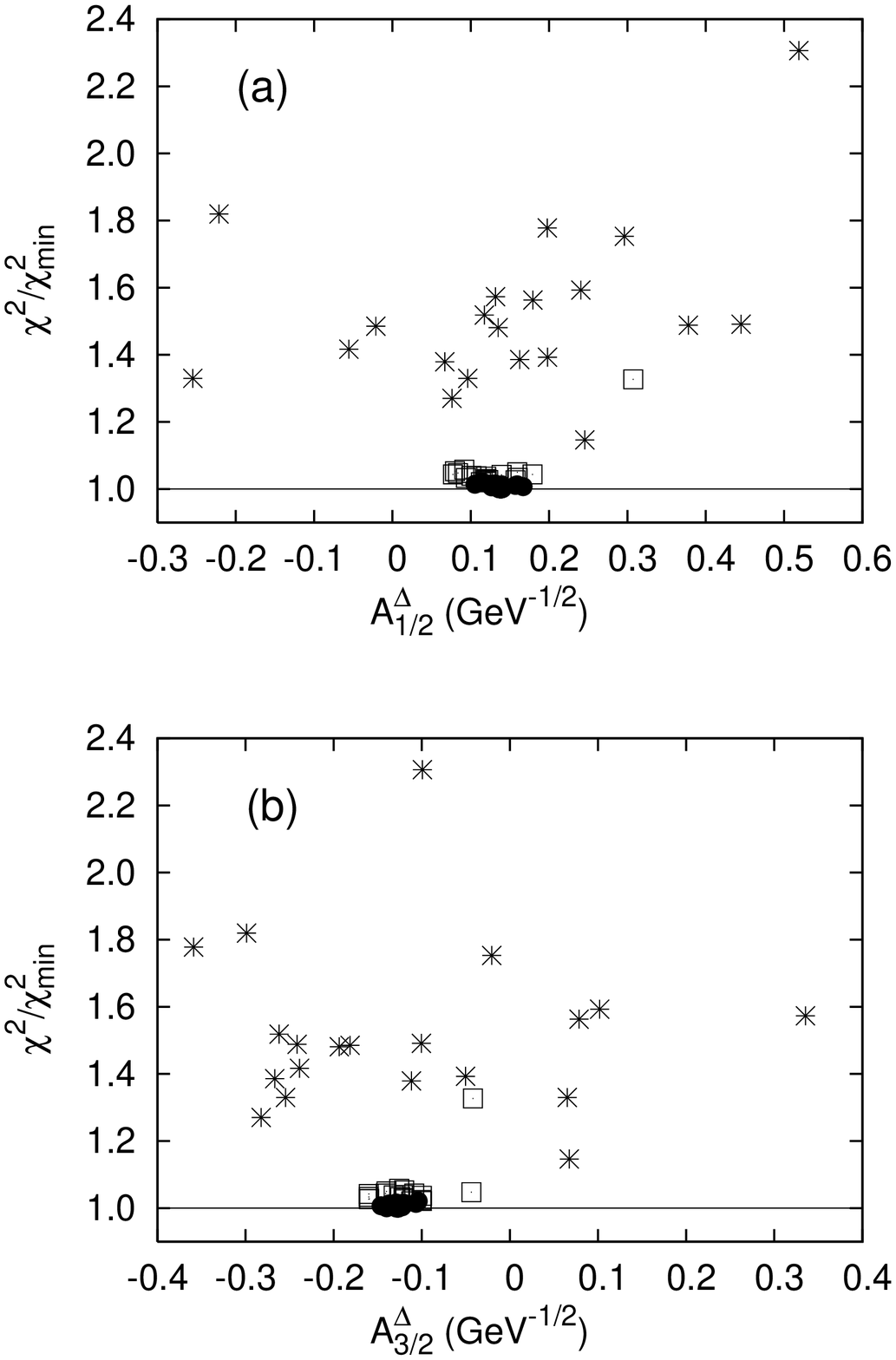}}}
\caption{Evolution of the minimization for the
helicity amplitudes of the $\Delta$(1700). 
Asterisks: minima obtained after 50 generations plus NAG;
open squares: minima obtained after 150 generations plus NAG;
solid circles: minima obtained after 400 generations plus NAG.} 
\label{fig:ev_delta1700}
\end{figure}

\begin{figure}
\rotatebox{-90}{\scalebox{0.3}[0.3]{\includegraphics{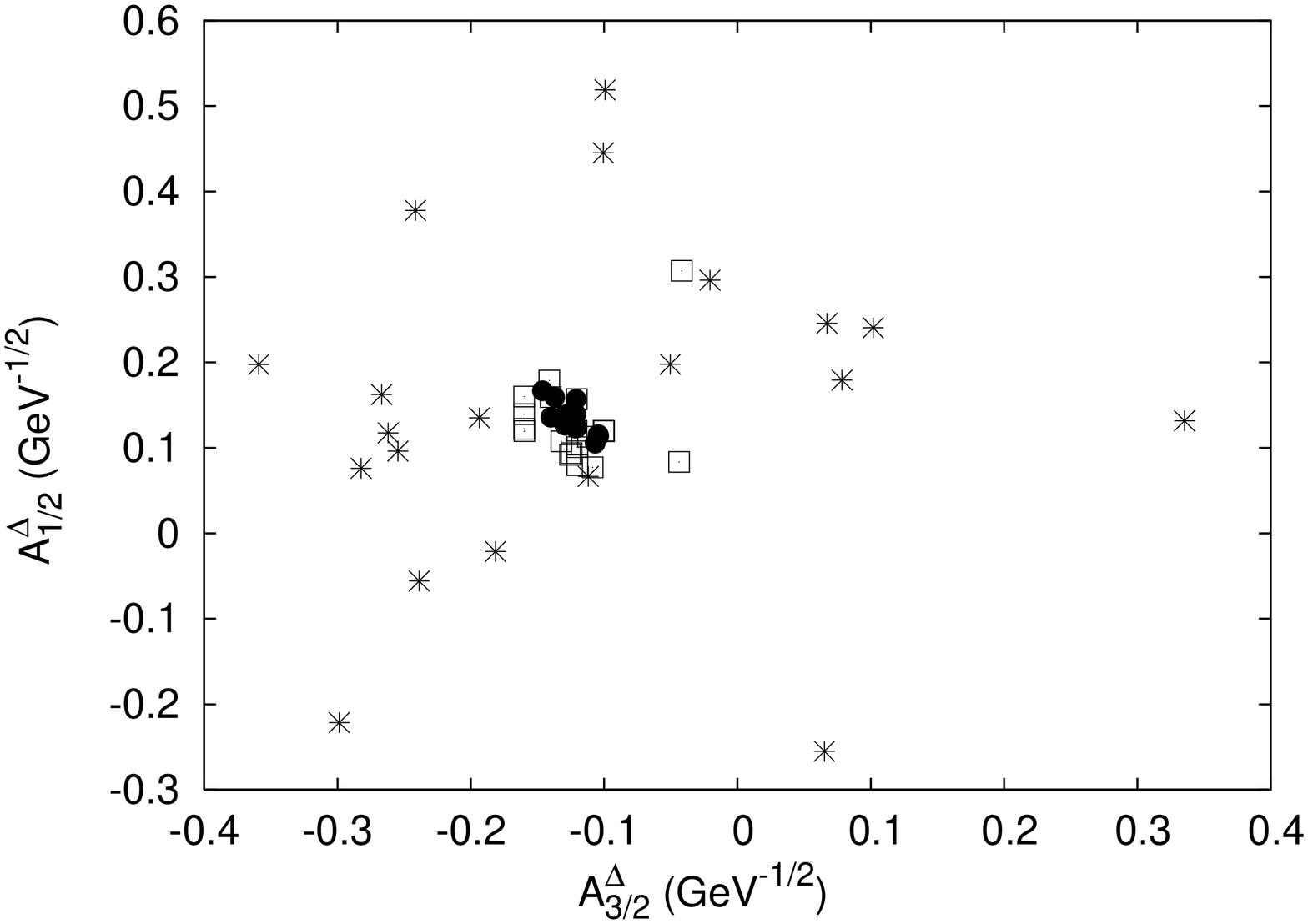}}}
\caption{Evolution of the minimization of the $\Delta$(1700)
helicity amplitudes. 
Same conventions as in Fig. \ref{fig:ev_delta1700}.}\label{fig:delta1700_ev}
\end{figure}

Figs. \ref{fig:delta1232} and \ref{fig:delta1700} show a typical situation that may arise when the parameters
are being determined. For $\Delta$(1232) the minimum is well-established and all the minima are constrained in a 
small region. The size of the region where the minima lie may provide a better estimation of the error 
associated with the parameters than the one provided by the correlation matrix. On the other hand, in 
Fig. \ref{fig:delta1232} the value for the $A^\Delta_{1/2}$ helicity amplitude appears to be in one of two 
split regions that are too close to be physically distinguished (left-upper panel). One region is centered at $-0.120$ 
and the other at $-0.119$ GeV$^{-1/2}$. The identification of these regions is one of the functionalities 
that GAs provide and one of their main advantages. When multiple regions containing minima of similar quality 
appear, the possible physical implications should be considered and further analysis to assess whether these 
different regions hold physical meaning (see subsection \ref{sec:oldfit}) is required.

We also show the minima of the $\Delta$(1700) that are constrained in just one region.
However, this region is larger than for the $\Delta$(1232) and the experimental information available for this 
resonance, thus, yields parameters that are not as well established as for other nucleon excitations.

The evolution of the position of the parameters for different instances of the GA$+$NAG procedure as the number 
of generations employed in the GA increases, is shown for the $\Delta$(1700) resonance in 
Figs. \ref{fig:ev_delta1700} and \ref{fig:delta1700_ev}. We can observe how the region of the minimum decreases 
while the GA evolves. For the 50 generations plus NAG run (asterisks in Figs. \ref{fig:ev_delta1700} 
and \ref{fig:delta1700_ev}) the $\chi^2$ is far away from its best value. 
This case exemplifies what happens when the parameters are assessed using the \texttt{E04FCF} after
the GA had not been converged and therefore the GA is merely providing 'very smart guesses', for the 
starting point of the gradient based routine. We observe in this case that the results spread over a wide 
range of values of the parameters, showing that indeed this is a hard optimization problem. Indeed, we expect 
that the starting values provided by unconverged instances of the GA are in fact much better  
that the ones we may figure without the aid of the GA. It is clear that to reach even an average quality 
optimum would be extremely hard (if not impossible) without the GA phase of our algorithm.
 After 150 generations plus NAG (open squares in 
Figs. \ref{fig:ev_delta1700} and \ref{fig:delta1700_ev}) the result looks much better, showing a region 
where the values of the parameters are well delimited. The $\chi^2$ is remarkably better and close to the 
best values obtained after 400 generations plus NAG (solid circles in 
Figs. \ref{fig:delta1700}, \ref{fig:ev_delta1700}, and \ref{fig:delta1700_ev}).

\begin{figure*}
\rotatebox{-90}{\scalebox{0.55}[0.55]{\includegraphics{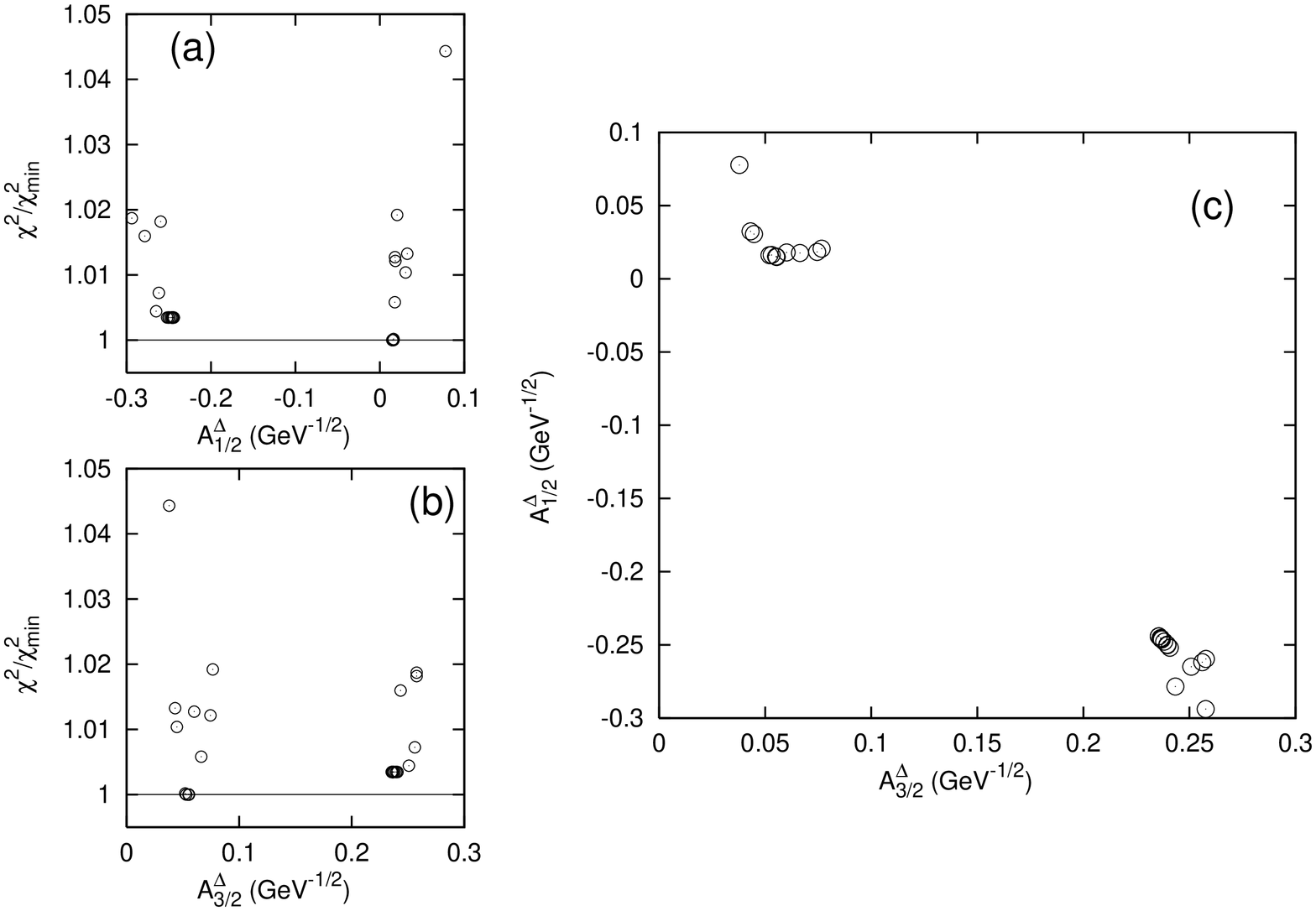}}}
\caption{Helicity amplitudes (equivalent to the coupling constants of the Lagrangians) of 
the $\Delta$(1700). We show thirty minima obtained in the
full minimization procedure (GA$+$NAG) for the 2005 SAID database
up to 1 GeV of photon energy with the model in Ref. \cite{fernandez06a}.
The upper hand left figure (a) shows the
$\chi^2/\chi^2_{\text{min}}$ versus the amplitude $A_{1/2}^\Delta$. The
lower left hand figure (b) shows the
$\chi^2/\chi^2_{\text{min}}$ versus the amplitude $A_{1/2}^\Delta$.
The figure on the right (c) shows $A^\Delta_{1/2}$ versus $A^\Delta_{3/2}$.} 
\label{fig:delta1700old}
\end{figure*}

\subsection{Minima Split in Various Regions} \label{sec:oldfit}

\begin{figure}
\rotatebox{-90}{\scalebox{0.3}[0.3]{\includegraphics{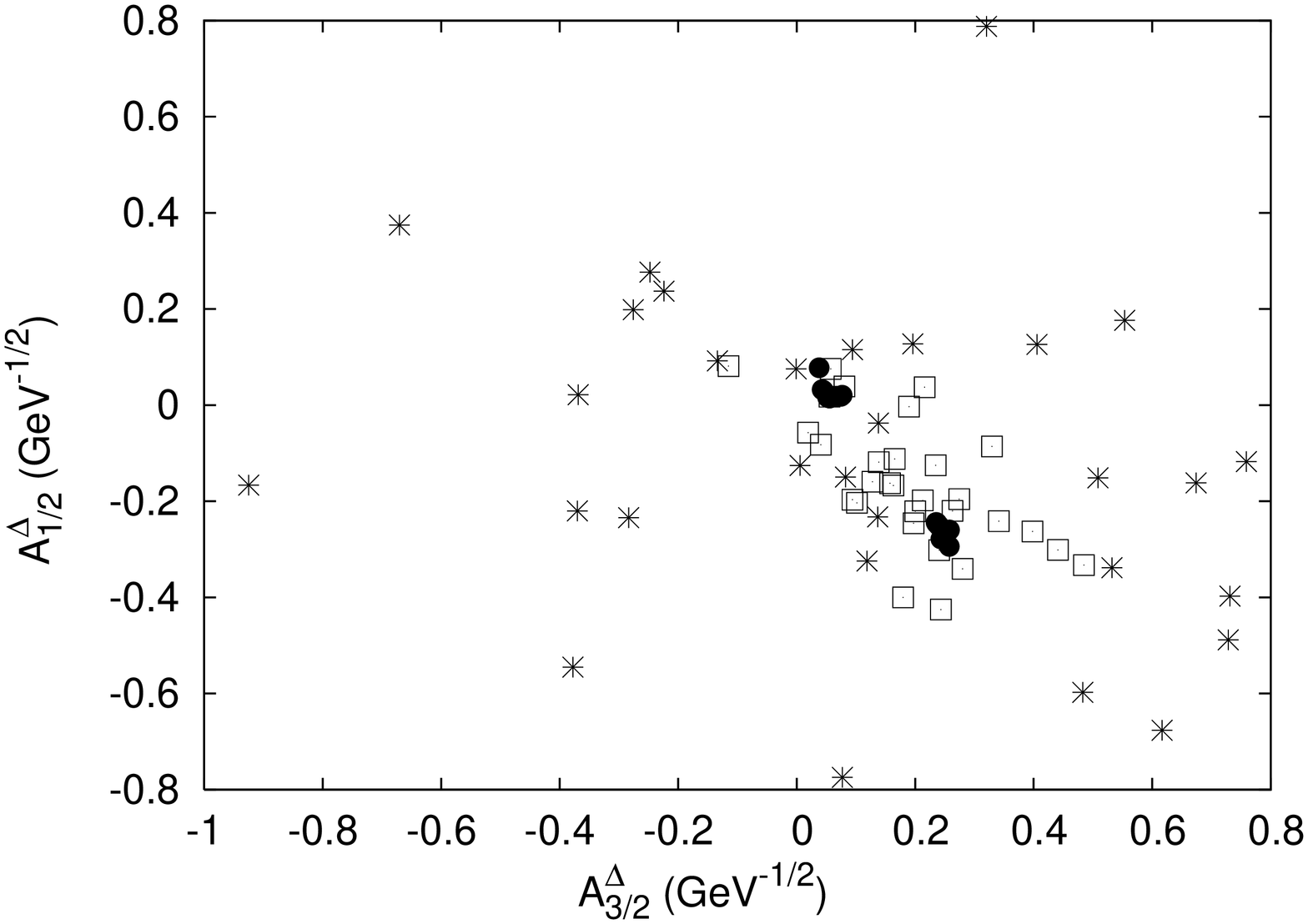}}}
\caption{Evolution of the minimization for the helicity amplitudes of the $\Delta$(1700) for the 2005 SAID 
database. Asterisks: minima obtained after 50 generations plus NAG; open squares: minima obtained after 
100 generations plus NAG; solid circles: minima obtained after 400 generations plus NAG.
The formation of the two regions where the minima group can clearly be seen.}
\label{fig:delta1700_evold}
\end{figure}

The amount and quality of data is of great importance in assessing  the parameters of any model. 
The pion photoproduction multipole data set employed for the fits in this work is the largest and of the highest 
quality ever available. It was released in 2006 and includes 3,760 data points.
It is interesting to see what would happen if we employ the 2005 SAID database instead, which includes 
up to 1.0 GeV photon energy  and considerably fewer (1526) data points, as done in 
Refs. \cite{phd,fernandez06a,fernandez06b}. We find that the results change for the not so well-determined 
resonances as is the case of the $\Delta$(1700) one. We find in this case several minima lying in more 
than one region. Fig. \ref{fig:delta1700old} is equivalent to Fig. \ref{fig:delta1700} but in this case 
fitting to the former data set. It becomes apparent how the minima split into two distinct regions. 
Fig. \ref{fig:delta1700_evold} is equivalent to Fig. \ref{fig:delta1700_ev} and shows how the regions are 
formed as the algorithm evolves. It also shows that a gradient method alone leads the optimization to incorrect
answers most of the time. There are several possible reasons for the appearance of this minima
structure. For instance, this can be caused by deficiencies in the model or in the data. We must keep 
in mind however, that this result can even have a physical meaning such as a possible shape coexistence for a state 
that can fit the data equally well for two sets of parameter values. This would have to be studied 
within a model in which the resonance is included as a combination of both states and re-fit to experimental data. 
However, it seems that this is not the situation we encountered here. The results presented in the previous 
subsection and in Fig.  \ref{fig:delta1700} clearly indicate that improving the database and extending  
the model to higher energies (which allows one to account for the tail of the $\Delta$(1700) 
resonance) are sufficient to collapse the two $\chi^2$ regions into one single region.

\section{Final remarks}\label{sec:conclusions}

We have presented a hybrid optimization procedure which combines a GA with the gradient-based 
routine \texttt{E04FCF} from the NAG libraries. We have successfully applied  this algorithm to determine 
the coupling constants of the low-lying nucleon resonances within a realistic Lagrangian model of the pion 
photoproduction reaction. The results for the couplings were summarized 
in Table \ref{tab:helicities}.

Traditional optimization tools are often useless for this kind of multi-parameter optimizations when the 
parameter space is large and the function to fit presents many local minima. The assessment 
of the low-lying resonances properties by means of reaction models is an example of a very 
difficult optimization problem for conventional algorithms \cite{Ireland,phd,fernandez06a}.

The hybrid optimization procedure presented in this paper is a powerful and versatile 
optimization tool that can be applied to many problems in physics that involve the 
determination of a set of parameters from data. It is a promising method for extracting 
both reliable physical parameters as well as their confidence intervals. Indeed, computing
correlations among different parameters by comparing different solutions obtained 
by the hybrid optimization method, in a manner similar to what is shown in the panel on the right 
in Fig. 6, is  probably more meaningful than the simple covariance matrices returned by gradient based 
optimization routines.

Finally, we have shown how we can use the procedure we have outlined to identify weaknesses in the model and 
assess the reliability of the parameters obtained. Not only the error bars 
have to be considered when quoting the uncertainty in the determination of a 
parameter, but also whether the minima are concentrated into one single region or 
split into several ones, and the possible physical explanations of such situation.

\begin{acknowledgments}
This work has been supported in part under contracts
FIS2005-00640, FPA2006-07393, and FPA2007-62216
of the Ministerio de Educaci\'on y Ciencia (Spain)
and by UCM and the Comunidad de Madrid
under project number 910059 (Grupo de F\'{\i}sica Nuclear).
C.F.-R. is supported by "Programa de becas posdoctorales"
of the Ministerio de Educaci\'on y Ciencia (Spain).
Part of the computations for this work were carried out at the 
``Cluster de C\'alculo de Alta Capacidad para T\'ecnicas F\'{\i}sicas'',
which is partly funded by the EU Commission under FEDER program and by
the Universidad Complutense de Madrid (Spain).
\end{acknowledgments}

\end{document}